\documentclass[preprintnumbers,twocolumn,article,amsmath,amssymb,floatfix,9pt,prd,superscriptaddress,nofootinbib]{revtex4-1}
\usepackage[utf8,latin1]{inputenc}
\usepackage{graphicx}
\usepackage{dcolumn}
\usepackage[dvipsnames]{xcolor}
\usepackage[T1]{fontenc}

\usepackage{mathrsfs}  
\usepackage{cases}
\usepackage{bm}
\usepackage{academicons}
\usepackage{mathtools, nccmath}
\usepackage{fancyhdr}
\usepackage{tikz,xcolor}

\usepackage{tensor}
\usepackage[normalem]{ulem}
\usepackage{lipsum}
\usepackage{soul}
\usepackage{cancel}
\usepackage{stackengine,scalerel}
\usepackage{mathabx}
\usepackage{hyperref}
\usepackage{tabularx}
\usepackage{makecell}
\hypersetup{colorlinks, linkcolor={red},citecolor={blue},urlcolor={blue}}  

\AtBeginDocument{\fontsize{8}{11}\selectfont} 

\definecolor{lime}{HTML}{A6CE39}
\DeclareRobustCommand{\orcidicon}{
	\begin{tikzpicture}
	\draw[lime, fill=lime] (0,0) 
	circle [radius=0.16] 
	node[white] {{\fontfamily{qag}\selectfont \tiny ID}};
	\draw[white, fill=white] (-0.0625,0.095) 
	circle [radius=0.007];
	\end{tikzpicture}
	\hspace{-2mm}
}

\fancypagestyle{plain}{%
  \fancyhf{}
  \fancyfoot[C]{\iffloatpage{}{\thepage}}
  }
\foreach \x in {A, ..., Z}{%
	\expandafter\xdef\csname orcid\x\endcsname{\noexpand\href{https://orcid.org/\csname orcidauthor\x\endcsname}{\noexpand\orcidicon}}
}

\begin{document}

\title[Holographic dark energy as a source for slowly rotating wormholes: Implications for null geodesics and shadows]{Holographic dark energy as a source for slowly rotating wormholes:\\ Implications for null geodesics and shadows}

\author{A. Errehymy\orcidA{}}%
\email{abdelghani.errehymy@gmail.com (Corresponding author)}
\affiliation{Astrophysics Research Centre, School of Mathematics, Statistics and Computer Science, University of KwaZulu-Natal, Private Bag X54001, Durban 4000, South Africa}
\affiliation{Center for Theoretical Physics, Khazar University, 41 Mehseti Str., Baku, AZ1096, Azerbaijan}

\author{S. K. Maurya \orcidB{}}
\email{sunil@unizwa.edu.om}
\affiliation{Department of Mathematical and Physical Sciences,\\ College of Arts and Sciences, University of Nizwa, Nizwa 616, Sultanate of Oman}

\author{M. Govender\orcidC{}}%
\email{megandhreng@dut.ac.za}
\affiliation{Department of Mathematics, Durban University of Technology, Durban 4000, South Africa}

\author{K. N. Singh \orcidD{}}
\email{ntnphy@gmail.com, nsingh@physics.du.ac.in}
\affiliation{Department of Physics \& Astrophysics, University of Delhi, Delhi 110007, India}

\author{J. Rayimbaev\orcidF{}}
\email{javlonrayimbaev6@gmail.com}
\affiliation{Institute of Theoretical Physics, National University of Uzbekistan, University Street 4, Tashkent 100174, Uzbekistan}
\affiliation{University of Tashkent for Applied Sciences, Str. Gavhar 1, Tashkent 100149, Uzbekistan}
\affiliation{Kimyo International University in Tashkent, Shota Rustaveli street 156, Tashkent 100121, Uzbekistan}

\author{B. Myrzakulova\orcidE{}}
\email{bz.myrzakulova@gmail.com}
\affiliation{Department of General and Theoretical Physics, L.N. Gumilyov Eurasian National University, Astana 010008, Kazakhstan }

\author{S. Murodov\orcidI{}}
\email{mursardor@ifar.uz}
\affiliation{New Uzbekistan University, Movarounnahr Str. 1, Tashkent 100000, Uzbekistan}
\affiliation{Institute of Fundamental and Applied Research, National Research University TIIAME, Kori Niyoziy 39, Tashkent 100000, Uzbekistan}

\date{\today} 

\begin{abstract}
{\footnotesize In this work, we explore for the first time slowly rotating traversable wormholes embedded in holographic dark energy. We focus on three representative holographic dark energy models---R\'{e}nyi, mixed, and Moradpour---and construct the wormhole shape functions directly from these energy density profiles using a Teo-type rotating wormhole metric. This allows us to examine the wormhole geometry in detail, including throat structure, the flaring-out condition for safe traversal, and violations of the null energy condition. To capture the effects of different redshift behaviors, we consider three smooth hyperbolic redshift functions---Sinh, Cosh, and Tanh---and study how they influence photon motion, null geodesics, effective potentials, photon-sphere locations, and Lense-Thirring precession caused by wormhole rotation. Our analysis shows that cuspy R\'{e}nyi profiles produce tighter photon orbits and stronger asymmetry, while smoother mixed and Moradpour profiles allow more circular paths and weaker frame-dragging effects. Finally, we calculate the shadows cast by these wormholes, finding that R\'{e}nyi-supported wormholes generate smaller, asymmetric shadows, whereas mixed and Moradpour-supported wormholes produce larger, nearly circular silhouettes. Altogether, this study provides a detailed theoretical picture of photon dynamics, shadow morphology, and relativistic effects in slowly rotating wormholes within realistic holographic dark energy environments, offering potential guidance for observational signatures of these exotic objects.
}
\end{abstract}

\maketitle
\section{Introduction}

Back in 1935, Einstein and Rosen investigated a way to describe matter and electricity without encountering singularities \cite{Einstein:1935tc}. Their approach imagined physical space as two identical layers linked by a connecting bridge. In this picture, particles themselves were the bridges joining the layers. By combining the general relativistic metric tensor \(g_{\mu\nu}\) with the electromagnetic fields \(\phi_\mu\), they altered the gravitational equations and showed that smooth, spherically symmetric solutions could exist. Today, these structures are famously known as Einstein-Rosen bridges---or, more commonly, wormholes.

The idea of viewing particles as bridges might have emerged as early as 1916. As Gibbons notes \cite{Gibbons:2015ed}, just a few months after Schwarzschild unveiled his solutions \cite{Schwarzschild:1916uq, Schwarzschild:1916ae}, Flamm submitted a manuscript exploring the geometric features of both the interior and exterior Schwarzschild spacetime \cite{Flamm:2015ogy}. For the exterior solution, Flamm showed that ``the planar section is isometric to a surface of revolution, where the meridional curve is a parabola.'' Yet, he never interpreted this geometry as a bridge connecting distinct regions of spacetime. Because of this, it is generally accepted that the work of Einstein and Rosen laid the foundation for the modern concept of wormholes.

At first, the concept of bridges captured more interest than black holes---until Fuller and Wheeler \cite{Fuller:1962zza} demonstrated that the Einstein-Rosen bridge\footnote{Also known as the Schwarzschild wormhole.} is inherently unstable. Their analysis revealed that such a bridge would pinch off in a finite time , making Schwarzschild wormholes non-traversable. Despite this limitation, the idea of traversable wormholes resurfaced in later studies \cite{Morris:1988cz, Morris:1988tu, Visser:1989kh, Visser:1989kg, Poisson:1989zz, Frolov:1990si, Visser:1990wi}. Notably, Morris and Thorne \cite{Morris:1988cz} introduced a new approach that laid out a set of basic wormhole criteria. By assuming a wormhole geometry and applying the Einstein field equations, they calculated key physical quantities, including energy density, tension per unit area, and pressure, expressed in terms of redshift and shape functions. Their framework identifies three essential features for a wormhole: the spacetime must be static and spherically symmetric; the solution must include a throat connecting two asymptotically flat regions, a property tied to the shape function; and the wormhole must be horizonless, ensuring causal connection between the two regions---achieved by keeping the redshift function finite everywhere.

Another key outcome of the Morris-Thorne analysis is the requirement of so-called exotic matter for sustaining a traversable wormhole. They argued that an observer moving through the throat at nearly light speed would measure a negative energy density. From a classical standpoint, this implies a violation of the standard energy conditions, which were formulated precisely to exclude such negative energies. Yet, quantum theory allows for circumstances where these conditions can be locally violated---particle creation processes provide one example \cite{Martin-Moruno:2013sfa}. For this reason, the existence of the unusual matter needed to keep a wormhole throat open cannot be completely dismissed.

Wormholes have been explored in a wide range of theoretical scenarios, from standard general relativity \cite{Echeverria:1991nk, Hochberg:1997wp, Willenborg:2018zsv, Willenborg:2018zsv, Benavides-Gallego:2021lqn} to various modified gravity frameworks \cite{Moffat:1991xp, Bhawal:1992sz, Letelier:1993cj, Vollick:1998qf, Myrzakulov:2015kda}. For instance, Benavides-Gallego and collaborators \cite{Benavides-Gallego:2021lqn} studied how spinning test particles move in the vicinity of a traversable wormhole. By employing the Mathisson-Papapetrou-Dixon equations, they derived the corresponding effective potential and found that it depends explicitly on the particle's dimensionless spin parameter \(s\). In the special case \(s=0\), the motion reduces to geodesic behavior, and the effective potential becomes perfectly symmetric, meaning that the particle experiences identical dynamics in both regions connected by the wormhole. As a consequence, the innermost stable circular orbit (ISCO) lies at the same radial distance from the throat on either side. When the particle's spin \(s\) is nonzero, the effective potential shows a symmetry: flipping both the spin and angular momentum (\(s,L \to -s,-L\)) leaves the dynamics unchanged, so a particle spinning along the axis and moving counterclockwise behaves like one with opposite spin moving clockwise \cite{Benavides-Gallego:2021lqn}. Considering both possibilities produces a mirrored behavior between the two universes, effectively matching the nonspinning case for equal angular momentum and opposite spin. This symmetry also affects the ISCO: nonspinning particles have a single symmetric orbit, while spinning particles have one ISCO if \(|s|\ge1\) and two if \(-1<s<1\), with one always closer to the throat. Since the momentum \(p^\alpha\) and velocity \(u^\alpha\) are not aligned, approaching the center can push \(u^\alpha u_\alpha\) toward divergence, leading to unphysical spacelike motion; enforcing \(u^\alpha u_\alpha \le 0\) keeps trajectories timelike, which for nonrotating wormholes requires \(|s|<1.5\).

Another promising direction explores the connection between wormhole physics and holographic dark energy (HDE). Building on the Bekenstein-Hawking relation between entropy and horizon area \cite{Bekenstein:1973ur, Hawking:1975vcx, Cohen:1998zx}, HDE models have been extended to include quantum corrections, often through R\'{e}nyi entropy \cite{Tsallis:1987eu, Moradpour:2017shy}. These modifications adjust the energy density in a way that improves thermodynamic consistency while also offering a mechanism for cosmic acceleration. Importantly, they can provide the exotic matter needed to support stable wormholes. In this context, Nashed and Eid \cite{Nashed:2026rah} have shown that HDE can act as a source for wormholes in modified gravity, highlighting its role in sustaining these structures. Earlier studies along similar lines include Chaudhary et al. \cite{Chaudhary:2025wsd}, who analyzed multiple HDE density profiles in modified gravity, and Paul and collaborators \cite{Paul:2025vem}, who confirmed that such profiles can support stable wormhole solutions.

In this work, we explore for the first time how slowly rotating traversable wormholes behave when immersed in HDE, focusing on three representative models: R\'{e}nyi, mixed, and Moradpour. Using a Teo-type rotating wormhole metric, we construct the shape functions directly from these HDE profiles, which lets us examine the wormhole geometry in detail. We study the throat structure, the flaring-out condition needed for safe traversal, and the associated violations of the null energy condition (NEC), all crucial for understanding whether such configurations are physically plausible. Beyond geometry, we track photon motion by analyzing null geodesics, effective potentials, photon-sphere locations, and Lense-Thirring (LT) precession caused by wormhole rotation. This approach shows how cuspy R\'{e}nyi profiles versus smoother mixed or Moradpour profiles affect photon paths, orbit asymmetry, and frame-dragging (FD) effects. Finally, we calculate the shadows cast by these wormholes, linking theory to potential observations. R\'{e}nyi-supported wormholes produce tighter, asymmetric shadows, while mixed and Moradpour-supported wormholes yield broader, nearly circular silhouettes. By combining wormhole geometry, energy condition violations, photon dynamics, and shadow features, this study provides, a novel, comprehensive framework for understanding slowly rotating wormholes in realistic HDE environments and their potential observable signatures.

The paper is organized as follows. In Sect. \ref{Sec:II}, we describe the geometry of stationary axisymmetric rotating wormholes. Sect. \ref{Sec:III} introduces common HDE density distributions, including: \(A-\)the R\'{e}nyi entropy-constrained model, \(B-\)the mixed energy density framework, and \(C-\)the Moradpour entropy-constrained model. In Sect. \ref{Sec:IV}, we construct and analyze the corresponding wormhole shape functions for these HDE models. Sect. \ref{Sec:V} is devoted to photon dynamics around slowly rotating HDE-supported wormholes, covering: \(A-\)FD and LT effects, \(B-\)photon motion and effective potentials, \(C-\)photon spheres and rotational splitting, and \(D-\)photon-sphere shifts induced by hyperbolic redshift profiles. Finally, Sect. \ref{Sec:VI} presents the results and discussion.

\section{Geometry of Stationary Axisymmetric Rotating Wormholes}\label{Sec:II}
We analyze the spacetime structure of a rotating wormhole that possesses both stationarity and axial symmetry. These symmetries guarantee the presence of two independent Killing vector fields \cite{Teo:1998dp},
\begin{eqnarray}
\zeta^{a} &=& \left(\frac{\partial}{\partial t}\right)^{a},
\label{eq:Killing_t}\\
\psi^{a} &=& \left(\frac{\partial}{\partial \varphi}\right)^{a},
\label{eq:Killing_phi}
\end{eqnarray}
which reflect the invariance of the geometry under time translations and rotations about the azimuthal axis, respectively.

From a general geometrical standpoint, any stationary and axisymmetric spacetime admits a line element that can be written as \cite{Papapetrou:1966, Carter:1968a, Carter:1968b, Carter:1987}
\begin{eqnarray}
ds^{2} = g_{tt} \, dt^{2} + 2 g_{t\varphi} \, dt \, d\varphi 
+ g_{\varphi\varphi} \, d\varphi^{2} + g_{ij} \, dx^{i} dx^{j},
\label{eq:general_metric} 
\end{eqnarray}
where the indices $i,j$ denote the spatial coordinates orthogonal to the $(t,\varphi)$ subspace.

It has also been argued that introducing a conformal factor depending on time into the Morris--Thorne wormhole metric may reduce the degree of NEC violation. However, this approach causes the throat to expand uniformly in time, thereby breaking the stationary nature of the spacetime and making the configuration unsuitable for sustained traversability \cite{Teo:1998dp, Roman:1993, Kar:1994, Kar:1996}.

In this work, we adopt a rotating wormhole geometry written in standard spherical coordinates, following the framework proposed in~\cite{Teo:1998dp}. The corresponding line element is given by
\begin{equation}
ds^{2} = - e^{2\Phi(r)} dt^{2} 
+ \frac{dr^{2}}{1 - \dfrac{\epsilon(r)}{r}} 
+ r^{2} K(r)^{2} \left[ d\theta^{2} 
+ \sin^{2}\theta \, ( d\varphi - \omega(r) dt )^{2} \right],
\label{metric_general2}
\end{equation}
where the function $\Phi(r)$ represents the gravitational redshift and must remain finite throughout the spacetime in order to avoid the appearance of event horizons and preserve traversability~\cite{Butcher:2015sea}. The quantity $\epsilon(r)$ characterizes the wormhole shape and determines the spatial embedding of the geometry. 
The term $\omega(r)$ encodes the local angular velocity of the rotating wormhole. The radial coordinate $r$ reaches its minimum value at the throat $r_{th}$, defined through the condition $\epsilon(r_{th}) = r_{th}$. Close to this region, the spacetime must satisfy the usual flaring-out requirement~\cite{Morris:1988cz},
\begin{eqnarray}
\epsilon(r) - r \, \epsilon'(r) \geq 0,
\end{eqnarray}
with the prime denoting differentiation with respect to $r$. Moreover, asymptotic flatness demands that $\epsilon(r)/r \to 0$ as $r \to \infty$, while the inequality $\epsilon(r)/r < 1$ should hold everywhere so that horizons do not form.

The function $K(r)$ is assumed to be positive and monotonically increasing, and it determines the proper radial distance in the geometry. In the regime of slow rotation, it is common to approximate $K(r) \simeq 1$. A similar metric structure has appeared in the analysis of slowly rotating relativistic stars performed by Hartle~\cite{Hartle:1967ha, Hartle:1967he}. For the spacetime to approach flatness at large distances, the metric functions must obey
\begin{eqnarray}
\Phi(r) \to 0, \qquad K(r) \to 1, \qquad \omega(r) \to 0 \quad (r \to \infty).
\label{asymptotic_condition}
\end{eqnarray}
At large radii, the angular velocity is typically expanded as
\begin{eqnarray}
\omega(r) = \frac{2 J}{r^{3}} + \mathcal{O}(r^{-4}),
\end{eqnarray}
where $J$ denotes the total angular momentum of the wormhole~\cite{Teo:1998dp}. Within the slow-rotation (linear) approximation, the influence of $\omega(r)$ on both the NEC and the form of the shape function remains subleading.

\section{Common HDE Density Distributions} \label{Sec:III}
\label{sect3}
A growing body of literature suggests that the observed cosmic acceleration may be consistently interpreted within HDE scenarios grounded in generalized entropy formalisms. In these approaches, the dark energy density is not inserted ad hoc, but instead arises from modifications of the horizon entropy relation, as proposed in the R\'{e}nyi and Moradpour extensions of the standard Bekenstein---Hawking framework. By altering the entropy---area law, one obtains new expressions for the effective energy density associated with the cosmological horizon, which naturally lead to different evolutionary behaviors of the Universe at late times. Such constructions provide a thermodynamic underpinning for dark energy and offer a unified way to connect quantum gravitational corrections with large-scale cosmology. Depending on the chosen entropy prescription, the resulting holographic density profiles exhibit distinct functional forms that can affect the Hubble expansion rate, the effective equation of state, and the transition from decelerated to accelerated expansion. These models have been widely explored in the context of modified gravity and quantum-corrected cosmology, where they serve as viable alternatives to the conventional HDE scenario. In the following, we outline the most commonly used HDE density models inspired by the R\'{e}nyi, mixed, and Moradpour entropy constructions and briefly discuss their cosmological implications \cite{Bekenstein:1973ur, Hawking:1975vcx, Li:2004rb, Hu:2006ar, Myung:2004ch, Moradpour:2017fmq, Manoharan:2022qll}.

\subsection{1\textsuperscript{st} model: HDE density constrained by the R\'{e}nyi entropy}
Within the holographic principle framework \cite{Myung:2004ch, Hu:2006ar}, the R\'{e}nyi-inspired HDE density provides a thermodynamically motivated description of how dark energy is distributed across spacetime. This formulation is particularly useful for investigating the Universe's expansion history, large-scale structure, and the interplay between dark energy and cosmic evolution.  

The R\'{e}nyi HDE density is given by \cite{Mamon:2016wow}:
\begin{eqnarray} 
\label{rho_R}
\rho_{_{\rm R\acute{e}nyi}}(r) = \frac{\alpha}{\lambda r^4} \, \log \big( 1 + \pi \lambda r^2 \big),
\end{eqnarray}
where $\lambda$ is the R\'{e}nyi parameter and both $\alpha$ and $\lambda$ are taken to be positive. This profile naturally decays to zero as \( r \to \infty \) or when \( \lambda \to \infty \), while in the limit \( \lambda \to 0 \), it reduces to the standard Bekenstein-Hawking HDE density.  

The logarithmic dependence in Eq.~\eqref{rho_R} reflects a correction stemming from generalized gravitational thermodynamics, where the conventional Bekenstein-Hawking entropy is replaced by the R\'{e}nyi entropy. Such a modification effectively alters the energy density, providing a bridge between horizon thermodynamics and the cosmic evolution of dark energy.

\subsection{2\textsuperscript{nd} model: HDE density constrained by the mixed energy density framework}

The so-called mixed energy density provides a phenomenological description of the dark matter halo distribution, combining standard and entropy-inspired corrections \cite{Anderhalden:2012qt}. It plays a crucial role in cosmology and exotic spacetimes, such as wormholes, by contributing to the stabilization of the throat through the interaction of dark matter, dark energy, and exotic matter.

The density profile is given by:
\begin{eqnarray}
\label{rho_mixed}
\rho_{_{\rm Mixed}}(r) = \frac{3\alpha^{2}}{8\pi^{2}} \Bigg[ \frac{\pi}{r^2} - \pi^{2} \lambda \log\Bigg( 1 + \frac{1}{\pi \lambda r^2} \Bigg) \Bigg],
\end{eqnarray}
where $\alpha$ and $\lambda$ are positive parameters. This density naturally approaches zero as $r \to \infty$ or $\lambda \to \infty$, and in the limit $\lambda \to 0$, it reduces to the standard Bekenstein-Hawking HDE density.

The first term, $\pi/r^2$, reproduces the familiar inverse-square behavior of classical dark matter halos, which can help explain the flattening of galactic rotation curves. The logarithmic term acts as a correction associated with generalized entropy effects, modifying the energy distribution in a way that supports cosmic acceleration and can aid in stabilizing wormhole structures. This profile thus bridges standard dark matter modeling and exotic spacetime applications.

\subsection{3\textsuperscript{nd} model: HDE density constrained by the Moradpour entropy}
In this subsection, we consider the Moradpour-inspired HDE density profile \cite{Garattini:2023wgk}:
\begin{eqnarray}
\label{rho_Moradpour2}
\rho_{_{\rm Moradpour}}(r) = \frac{\alpha}{4 \pi r^2 (\pi \lambda r^2 + 1)},
\end{eqnarray}
where $\alpha$ and $\lambda$ are positive parameters.  

This formalism, motivated by R\'{e}nyi entropy modifications, has been applied to model dark matter density distributions in both astrophysical and cosmological contexts. The profile exhibits a naturally decaying behavior with radius $r$, consistent with the structure of galactic halos. In particular, Moradpour-inspired density profiles can be used to refine the standard Navarro-Frenk-White profile \cite{Moradpour:2018ivi}, introducing entropy-based corrections that influence the dark matter distribution. The parameter $\lambda$ encodes the deviation from the classical profile, providing a smooth decay of the energy density at large distances while maintaining consistency with observational data.


\begin{figure*}
\begin{center}
\includegraphics[width=18.5cm,height=10.0cm]{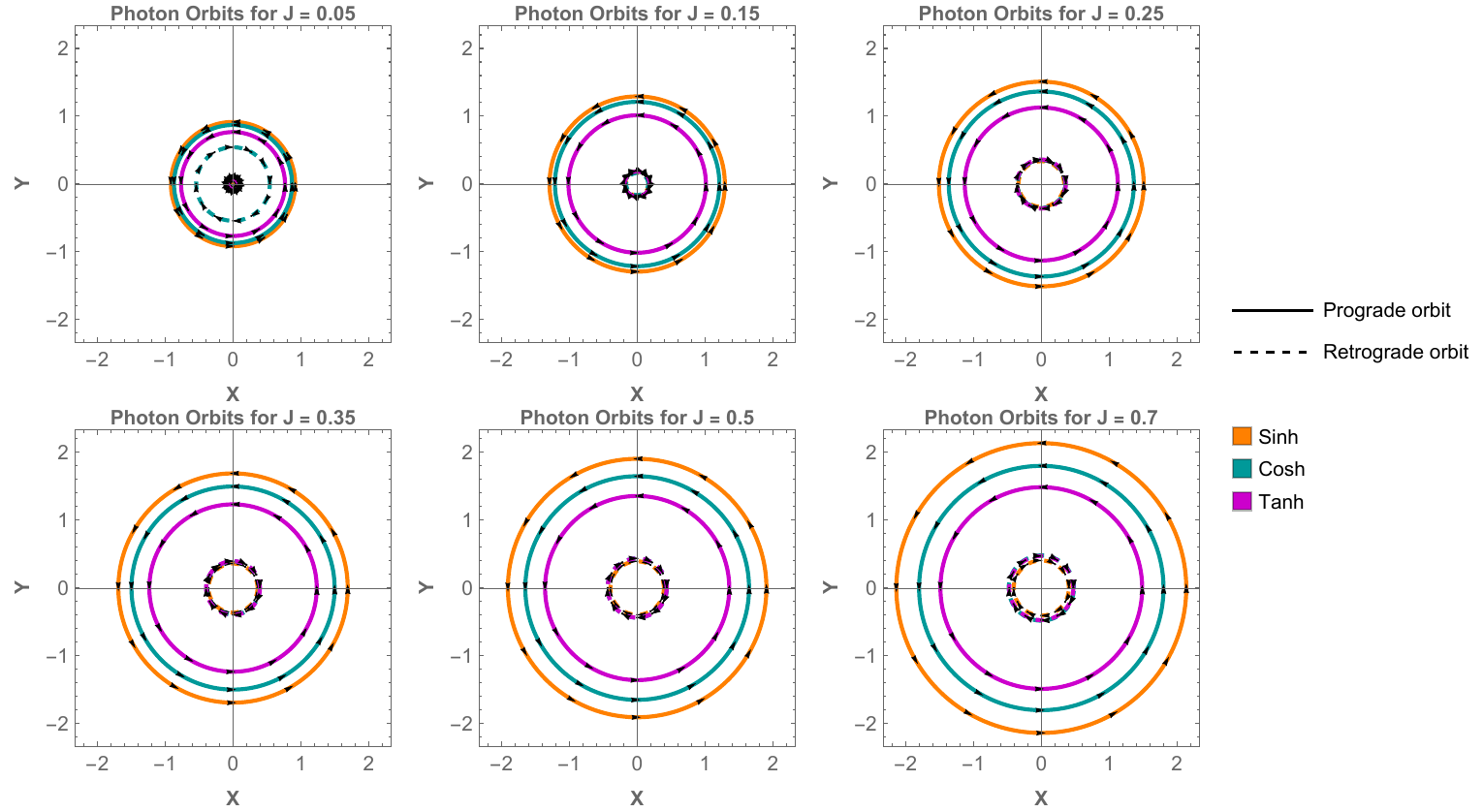}
\end{center}
\caption{{\footnotesize The figure presents photon orbits around rotating wormholes for six spin values, \(J = 0.05, 0.15, 0.25, 0.35, 0.5, 0.7\), arranged across six panels. Each panel compares three hyperbolic redshift profiles: Sinh (orange), Cosh (teal), and Tanh (magenta). Solid curves represent prograde photon orbits, moving in the same direction as the wormhole spin, while dashed curves represent retrograde photon orbits, moving in the opposite direction. Arrows along the curves indicate the direction of photon motion, providing a clear visual cue for orbit orientation. The figure demonstrates how increasing the wormhole spin amplifies the separation between prograde and retrograde photon orbits, a direct consequence of the FD effect. Comparing the three hyperbolic redshift profiles, one can see that the photon-sphere radii vary subtly with the chosen redshift function, highlighting the sensitivity of orbit locations to the wormhole's gravitational structure. These variations can influence the apparent shadow and lensing patterns of rotating wormholes, offering potential observational signatures. Overall, the figure captures both the dynamical behavior of photon orbits under spin and the effect of different redshift profiles on the photon-sphere geometry.}}\label{Fig1}
\end{figure*}

\section{Wormhole Shape Functions in HDE Models}\label{Sec:IV}

The shape function $\epsilon(r)$ is central to defining the geometry of static and slowly rotating wormholes, as it determines the curvature of spacetime around the throat. When the spacetime is spherically symmetric and sourced by a HDE density $\rho(r)$, the Einstein field equations lead to
\begin{equation}
\epsilon'(r) = 8 \pi r^2 \rho(r),
\end{equation}
where the prime denotes differentiation with respect to $r$, $\epsilon'(r) \equiv d\epsilon/dr$. The integration constant $\epsilon_{th}$ is fixed by the throat condition $\epsilon(r_{th}) = r_{th}$, ensuring the wormhole throat is well-defined and traversable.

The radial NEC takes the form
\begin{equation}
\rho + p_r = \frac{1}{8 \pi r^2} \left( \epsilon' - \frac{\epsilon}{r} \right) + \frac{1 - \epsilon/r}{4 \pi r} \Phi'(r),
\end{equation}
where $\Phi(r)$ represents the redshift function, controlling the gravitational redshift along the radial direction. By plugging in HDE-based shape functions and their derivatives, one can directly check where the NEC holds and where it is violated.

Studying the NEC in this context gives a clear picture of how HDE shapes the wormhole geometry. It highlights the conditions necessary for the throat to flare out properly and shows how the energy density influences the overall curvature of the spacetime. Here, we explore explicit wormhole geometries supported solely by HDE. By substituting the corresponding shape functions \(\epsilon(r)\) for three representative HDE models, we can see how the choice of dark energy profile shapes the wormhole throat, determines the flaring-out behavior, and sets the amount of exotic matter needed to sustain the structure. We also extend the discussion to slowly rotating configurations, where FD effects become relevant.

The general static, spherically symmetric wormhole metric is
\begin{equation}
ds^2 = - e^{2\Phi(r)} dt^2 + \frac{dr^2}{1 - \dfrac{\epsilon(r)}{r}} + r^2 \left(d\theta^2 + \sin^2\theta , d\varphi^2\right),
\end{equation}
with \(\epsilon(r)\) encoding the wormhole shape and \(\Phi(r)\) representing the redshift function.

The explicit form of the shape function depends on the chosen HDE density model:
\begin{itemize}
\item R\'{e}nyi-inspired HDE
   \begin{equation}
   \epsilon_{_{\rm R\acute{e}nyi}}(r) = \frac{8 \pi \alpha}{\lambda} \Big[2 \sqrt{\pi \lambda} \, \tan^{-1}\big(\sqrt{\pi \lambda} \, r\big)-\frac{\log(1 + \pi \lambda r^2)}{r}  \Big] + \epsilon_{\rm th}.
   \end{equation}
\item Mixed HDE
   \begin{equation}
   \epsilon_{_{\rm Mixed}}(r) = 3 \alpha^2 \Big[ r + \frac{r^2}{4} \log\Big(1 + \frac{1}{\pi \lambda r^2}\Big) + \frac{r}{2} - \frac{1}{4 \pi \lambda} \log\Big(1 + \frac{1}{\pi \lambda r^2}\Big) \Big] + \epsilon_{\rm th}.
   \end{equation}
\item Moradpour HDE
   \begin{equation}
   \epsilon_{_{\rm Moradpour}}(r) = \frac{2 \alpha}{\sqrt{\pi \lambda}} \, \arctan\big(\sqrt{\pi \lambda} \, r \big) + \epsilon_{\rm th}.
   \end{equation}
\end{itemize}
The constant \(\epsilon_{\rm th}\) ensures that the wormhole throat satisfies \(\epsilon_{\rm th} = r_{\rm th}\).

Introducing slow rotation modifies the spacetime by incorporating FD effects, without significantly altering the radial structure. The metric becomes
\begin{equation}
ds^2 = - e^{2\Phi(r)} dt^2 + \frac{dr^2}{1 - \epsilon(r)/r} + r^2 \left[d\theta^2 + \sin^2\theta \left(d\varphi - \omega(r) dt\right)^2\right],
\end{equation}

where \(\omega(r) \simeq 2 J / r^3\) is the angular velocity related to the total angular momentum \(J\). Each HDE profile leads to a distinct throat geometry, affecting the flaring-out condition and the distribution of exotic matter, while the spacetime remains asymptotically flat (\(\epsilon(r)/r \to 0\) as \(r \to \infty\)).

\section{Photon Dynamics Around Slowly Rotating HDE Wormholes}\label{Sec:V}
The motion of light near a wormhole is shaped by the spacetime geometry and by the wormhole's rotation. For wormholes supported by HDE, the radial profile is governed by the shape function \(\epsilon(r)\), while slow rotation introduces FD through the angular velocity \(\omega(r)\). In the equatorial plane (\(\theta = \pi/2\)), the line element reduces to
\begin{eqnarray}
ds^2 = - e^{2\Phi(r)} dt^2 + \frac{dr^2}{1 - \epsilon(r)/r} + r^2 (d\varphi - \omega(r) dt)^2.
\end{eqnarray}

We consider three smooth hyperbolic redshift functions to explore different HDE scenarios:
\begin{eqnarray}
\Phi_{\mathrm{Sinh}}(r) &=& \Phi_0 \, \sinh(r/r_0) \, e^{-r/r_c}, \label{Phi_1}\\
\Phi_{\mathrm{Cosh}}(r) &=& \Phi_0 \left[1 - \frac{1}{\cosh(r/r_0)}\right], \label{Phi_2}\\
\Phi_{\mathrm{Tanh}}(r) &=& \Phi_0 \, \tanh(r/r_0), \label{Phi_3}
\end{eqnarray}
all finite at the throat, asymptotically flat, and slightly distinct to illustrate different HDE effects on photon motion.

\subsection{FD and LT Effects}

Slow rotation introduces a LT precession, causing nearby photons to experience subtle directional bending. In the equatorial plane, the LT frequency is
\begin{equation}
\Omega_\mathrm{LT}(r) = - \frac{1}{2} \frac{d\omega(r)}{dr}.
\end{equation}

For typical slow rotation, \(\omega(r) \simeq 2J / r^3\), this gives
\begin{equation}
\Omega_\mathrm{LT}(r) = \frac{3J}{r^4},
\end{equation}
indicating that FD intensifies near the wormhole throat. Representative values near typical throat radii:
\begin{itemize}
\item \textbf{R\'{e}nyi:} $r_{th} = 0.2$, $J = 0.025$, $\Omega_\mathrm{LT}^\mathrm{R\acute{e}nyi} \approx 93.8$, (compact, cusp-like configuration)
\item \textbf{Mixed:} $r_{th} = 0.35$, $J = 0.05$, $\Omega_\mathrm{LT}^\mathrm{Mixed} \approx 10.5$, (intermediate core-cusp mix)
\item \textbf{Moradpour:} $r_{th} = 0.5$, $J = 0.075$, $\Omega_\mathrm{LT}^\mathrm{Moradpour} \approx 2.4$, (larger, smoother core)
\end{itemize}

\begin{figure*}
\begin{center}
\includegraphics[width=18.5cm,height=10.0cm]{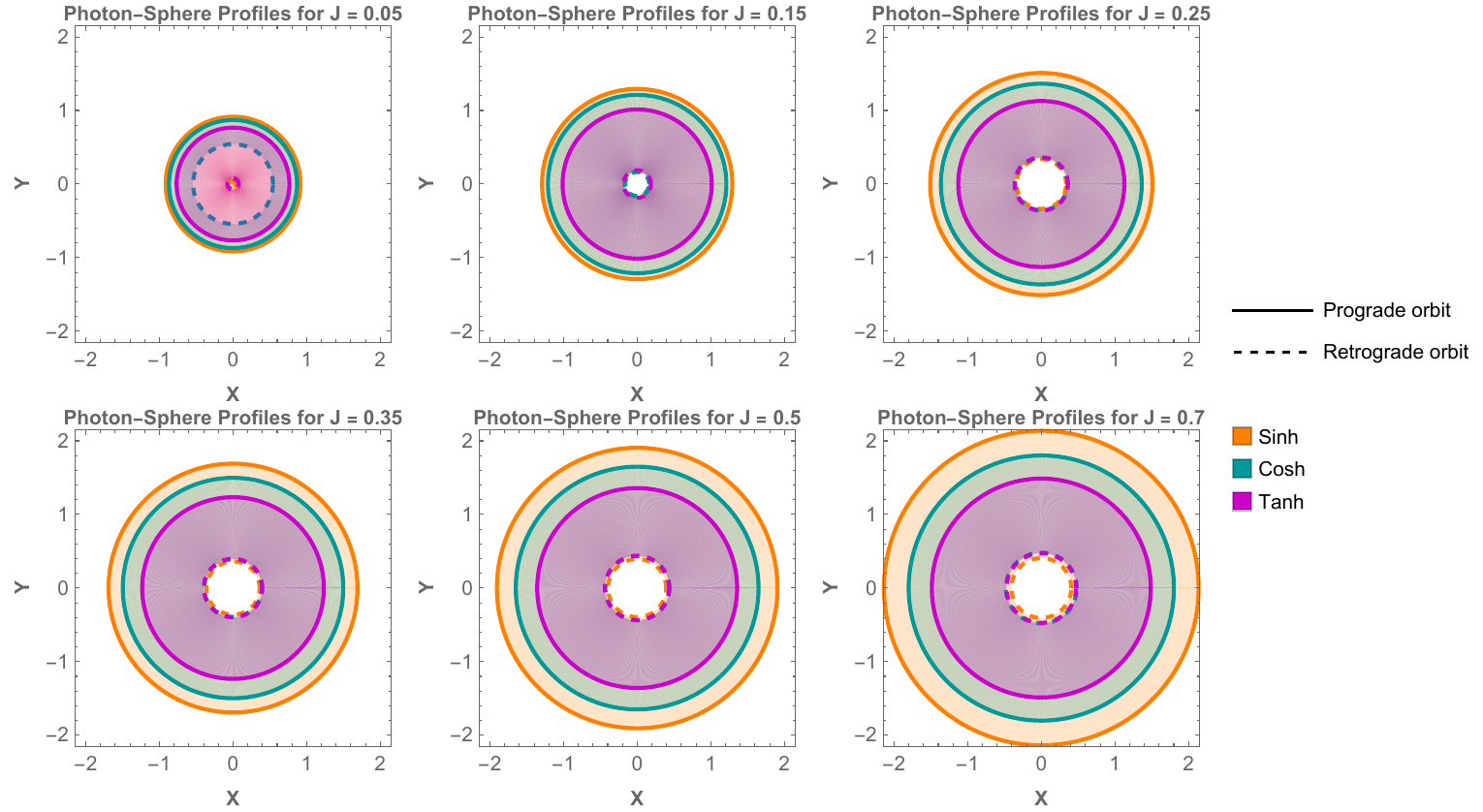}
\end{center}
\caption{{\footnotesize The figure shows photon-sphere profiles around rotating wormholes for six different spin values, \(J = 0.05, 0.15, 0.25, 0.35, 0.5, 0.7\), presented across six panels. Each panel compares three hyperbolic redshift profiles: Sinh (orange), Cosh (teal), and Tanh (magenta). Solid lines represent prograde photon-sphere orbits, moving in the same direction as the wormhole spin, while dashed lines represent retrograde orbits, moving in the opposite direction. The figure demonstrates how increasing spin causes the prograde and retrograde photon-sphere radii to diverge, illustrating the FD effect. The differences among the three hyperbolic redshift profiles show how the choice of redshift function affects the location of the photon spheres, indicating how these variations could influence the apparent shadows and lensing patterns of rotating wormholes. This comprehensive view captures both the dynamical impact of spin and the sensitivity of photon-sphere geometry to the wormhole's redshift profile, providing important insights into their observational signatures.}}\label{Fig2}
\end{figure*}

\subsection{Photon Motion and Effective Potential}

Photon trajectories respect conservation of energy \(E\) and angular momentum \(L\):
\begin{eqnarray}
E &=& e^{2\Phi(r)} \, \dot t + r^2 \, \omega(r) \, \dot\varphi, \\
L &=& r^2 \, (\dot\varphi - \omega(r) \, \dot t), \\
b &=& \frac{L}{E}, \\
\dot r^2 &=& \left(1 - \frac{\epsilon(r)}{r}\right) \frac{E^2}{e^{2\Phi(r)}} \Bigg[ \left(1 - \omega(r) b \, e^{\Phi(r)} \right)^2 \nonumber\\&&- \frac{b^2 e^{2\Phi(r)}}{r^2} \Bigg] \equiv -V(r).
\end{eqnarray}

where \(V(r)\) is the effective potential. Maxima of \(V(r)\) identify photon spheres, and photon paths follow \(\dot r^2 + V(r) = 0\).

The azimuthal evolution along the trajectory is
\begin{equation}
\frac{d\varphi}{dr} =
\frac{b + r^2 \, \omega(r) \, e^{-2\Phi(r)}}{\sqrt{r(r - \epsilon(r)) \left[ \frac{(1 - \omega(r) b)^2}{e^{2\Phi(r)}} - \frac{b^2}{r^2} \right]}} \, E.
\end{equation}

Expanding to linear order in \(\omega(r)\), the total bending angle becomes
\begin{equation}
\hat{\alpha} = 2 \int_{r_\mathrm{min}}^\infty
\frac{b + r^2 \, \omega(r) \, e^{-2\Phi(r)}}{\sqrt{r(r - \epsilon(r)) \left[ (1 - \omega(r) b)^2 e^{-2\Phi(r)} - \frac{b^2}{r^2} \right]}} \, dr - \pi.
\end{equation}

Co-rotating photons (\(b>0\)) are slightly less deflected, while counter-rotating photons (\(b<0\)) bend more due to the rotation-induced asymmetry.

\subsection{Photon Spheres and Rotational Splitting}

Circular photon orbits are fixed by the relation
\begin{equation}
(1 - \omega(r) b)^2 = \frac{b^2 \, e^{2\Phi(r)}}{r^2}.\label{bph}
\end{equation}

For such orbits, the impact parameters corresponding to co-rotating and counter-rotating photons can be approximated as
\begin{eqnarray}
b_\pm(r_\mathrm{ph}) \simeq r_\mathrm{ph} e^{-\Phi_\mathrm{ph}} \pm r_\mathrm{ph}^2 \, \omega_\mathrm{ph} \, e^{-2\Phi_\mathrm{ph}},
\end{eqnarray}
where \( \Phi_\mathrm{ph} \equiv \Phi(r_\mathrm{ph})\), and \(\omega_\mathrm{ph} \equiv \omega(r_\mathrm{ph})\). The plus and minus signs distinguish the counter-rotating and co-rotating trajectories, respectively. In the absence of rotation, \(\omega(r_\mathrm{ph}) \to 0\), both expressions reduce to the same static value
\begin{equation}
b_\mathrm{ph} = \frac{r_\mathrm{ph}}{e^{\Phi(r_\mathrm{ph})}}.
\end{equation}
This shows that the HDE background modifies the shadow in two distinct ways: through changes in the photon-sphere radius \(r_\mathrm{ph}\) and through the local FD term \(\omega(r_\mathrm{ph})\). Configurations with a steeper HDE profile generally lead to a larger separation between co-rotating and counter-rotating photon paths, while smoother profiles tend to diminish this difference. Such behavior could, in principle, provide observational clues about both the underlying HDE model and the rotation of the wormhole.

The apparent boundary of the shadow is set by the photon impact parameters evaluated at the photon-sphere radius \(r_\mathrm{ph}\), as defined in Eq.~(\ref{bph}). This relation makes it clear that both the overall shadow size and its rotational asymmetry are governed by the redshift function \(\Phi(r)\) together with the slow-rotation FD term \(\omega(r)\). When the redshift is modeled through hyperbolic HDE profiles---specifically the \(\sinh\), \(\cosh\), and \(\tanh\) forms---a systematic behavior emerges. The steeper \(\sinh\)-type profile enhances gravitational focusing near the throat, typically producing larger and more distorted shadows. By contrast, the smoother \(\cosh\) and \(\tanh\) profiles lead to more moderate and nearly circular shadow shapes. In the limiting case of a vanishing redshift function, the shadow becomes smaller and more symmetric, while its detailed appearance still depends on the observer's inclination and the emission properties of the surrounding matter. Figs.~\ref{Fig1} and \ref{Fig2} show how light moves around rotating wormholes, revealing the structure of photon-spheres and photon orbits for six spin values: (J = 0.05, 0.15, 0.25, 0.35, 0.5, 0.7). Each panel compares three different hyperbolic redshift profiles---Sinh (orange), Cosh (teal), and Tanh (magenta)---which change the gravitational potential experienced by photons and shift their paths accordingly. The solid lines represent prograde orbits, moving with the wormhole's rotation, while dashed lines show retrograde orbits, moving against it. As the spin increases, the prograde orbits are pushed outward and the retrograde orbits are pulled inward, clearly demonstrating the FD effect. The differences between the redshift profiles show how the wormhole's internal geometry influences photon motion, subtly modifying the size and shape of the photon-spheres and the trajectories of photon orbits. These variations would affect what a distant observer sees, altering the apparent shadow, the shape of the light rings, and the gravitational lensing patterns, and providing insight into how rotation and spacetime curvature work together to shape the paths of light around these exotic objects.

At the same time, rotation slightly shifts the photon-sphere radius itself, which can be expressed as
\begin{equation}
r_\mathrm{ph} \simeq r_\mathrm{ph}^{(c)} + \delta r_\mathrm{ph}, \quad
\delta r_\mathrm{ph} \simeq \frac{2 r_\mathrm{ph}^{(c)3} \omega(r_\mathrm{ph}^{(c)})}{2 - r_\mathrm{ph}^{(c)} \Phi'(r_\mathrm{ph}^{(c)})}.
\end{equation}
Thus, both the displacement of the photon sphere and the splitting of impact parameters work together to shape the observable shadow in rotating HDE-supported wormhole spacetimes. For the three hyperbolic HDE redshift functions given in Eqs.~(\ref{Phi_1}-\ref{Phi_3}), one obtains:

\begin{eqnarray}
\delta r_\mathrm{ph}^{(\mathrm{Sinh})} &\simeq&\ \frac{4 J}{2 - r_\mathrm{ph}^{(c)} \, \Phi_0 \, e^{-r_\mathrm{ph}^{(c)}/r_c} \left[ \frac{\cosh(r_\mathrm{ph}^{(c)}/r_\mathrm{th})}{r_\mathrm{th}} - \frac{\sinh(r_\mathrm{ph}^{(c)}/r_\mathrm{th})}{r_c} \right]}, ~~~~~\\
\delta r_\mathrm{ph}^{(\mathrm{Cosh})} &\simeq&\ \frac{4 J}{2 - r_\mathrm{ph}^{(c)} \, \Phi_0 \, \frac{\sinh(r_\mathrm{ph}^{(c)}/r_\mathrm{th})}{r_\mathrm{th} , \cosh^2(r_\mathrm{ph}^{(c)}/r_\mathrm{th})}}, \\
\delta r_\mathrm{ph}^{(\mathrm{Tanh})} &\simeq&\ \frac{4 J}{2 - r_\mathrm{ph}^{(c)} \, \Phi_0 \, \frac{1}{r_\mathrm{th}} \, \operatorname{sech}^2(r_\mathrm{ph}^{(c)}/r_\mathrm{th})}.
\end{eqnarray}

Steeper \(\sinh\)-type HDE produces stronger co-/counter-rotating splitting, while smoother \(\cosh\) and \(\tanh\) profiles lead to nearly symmetric photon spheres. These differences directly shape the wormhole shadow, offering potential observational signatures of both the HDE profile and slow rotation.

\subsection{Photon-sphere shifts induced by hyperbolic redshift profiles}

The hyperbolic redshift functions introduced in Eqs.~(\ref{Phi_1}-\ref{Phi_3}) slightly modify the position of the photon sphere relative to the throat value $r_{th}$ that would arise in the absence of a redshift factor. We parameterize this deviation as
\begin{eqnarray}
r_{\mathrm{ph}} = r_{th} + \delta r, \qquad |\delta r| \ll r_{th},
\end{eqnarray}
so that the correction remains perturbatively small. At leading order, the shift is controlled by the local slope of the redshift function together with the underlying wormhole geometry,
\begin{eqnarray}
\delta r \simeq -\frac{r_{th}^{3}}{2}\,\Phi'(r_{th})\Big(1-\epsilon'(r_{th})\Big)^{-1}.
\end{eqnarray}

For the three hyperbolic choices, the corresponding derivatives read
\begin{eqnarray}
\Phi'_{\mathrm{Sinh}}(r) &=& \Phi_0 e^{-r/r_c} \left[\frac{\cosh(r/r_0)}{r_0}-\frac{\sinh(r/r_0)}{r_c}\right],\\
\Phi'_{\mathrm{Cosh}}(r) &=& \frac{\Phi_0}{r_0} \, \operatorname{sech}\left(\frac{r}{r_0}\right)\tanh\left(\frac{r}{r_0}\right),\\
\Phi'_{\mathrm{Tanh}}(r) &=& \frac{\Phi_0}{r_0}\,\operatorname{sech}^{2}\left(\frac{r}{r_0}\right).
\end{eqnarray}

Substituting these expressions into the general formula yields the explicit photon-sphere shifts:
\begin{itemize}
\item \textbf{Sinh profile:}
\begin{eqnarray}
\delta r_{\mathrm{Sinh}} \simeq -\frac{\Phi_0 r_{th}^{3}}{2\left(1-\epsilon'(r_{th})\right)}
e^{-r_{th}/r_c}\left[\frac{\cosh(r_{th}/r_0)}{r_0}-\frac{\sinh(r_{th}/r_0)}{r_c}\right].~~~
\end{eqnarray}
Here, the exponential damping competes with the growing hyperbolic term, determining the net displacement close to the throat.

\item \textbf{Cosh profile:}
\begin{eqnarray}
\delta r_{\mathrm{Cosh}} \simeq -\frac{\Phi_0 r_{th}^{3}}{2 r_0\left(1-\epsilon'(r_{th})\right)}
\operatorname{sech}\left(\frac{r_{th}}{r_0}\right)\tanh\left(\frac{r_{th}}{r_0}\right).
\end{eqnarray}
Since the derivative remains bounded, the resulting shift varies smoothly and stays moderate in magnitude.

\item \textbf{Tanh profile:}
\begin{eqnarray}
\delta r_{\mathrm{Tanh}} \simeq -\frac{\Phi_0 r_{th}^{3}}{2 r_0\left(1-\epsilon'(r_{th})\right)}
\operatorname{sech}^{2}\left(\frac{r_{th}}{r_0}\right).
\end{eqnarray}
Because the $\tanh$ function quickly approaches saturation, the correction is tightly localized near the throat and leads to a well-controlled inward displacement.
\end{itemize}

In summary, the size of $\delta r$ is essentially dictated by the behavior of $\Phi'(r_{th})$. The $\sinh$ model can generate stronger variations depending on the damping scale $r_c$, the $\cosh$ profile produces a finite and intermediate response, while the $\tanh$ case yields the most localized and gently regulated shift of the photon sphere.

\section{Results and Discussion}\label{Sec:VI}

In this work, we explore for the first time, as we acknowledge the pioneering contributions of previous studies on traversable wormholes, the dynamics of slowly rotating wormholes embedded within HDE environments. We investigate how the HDE density distribution, redshift profile, and slow rotation collectively determine photon motion, photon-sphere locations, and the resulting shadow. To capture a broad range of effects, we consider three representative HDE models: R\'{e}nyi, mixed, and Moradpour, each of which shapes the wormhole geometry, throat flaring, and the requirements for exotic matter in a distinct manner.

Photon trajectories confined to the equatorial plane obey
\begin{eqnarray*}
    \dot r^2 = \left(1 - \frac{\epsilon(r)}{r}\right) \frac{E^2}{e^{2\Phi(r)}} \left[ \left(1 - \omega(r) L e^{\Phi(r)}/E \right)^2 - \frac{L^2 e^{2\Phi(r)}}{E^2 r^2} \right],
\end{eqnarray*}
where \(E\) and \(L\) are conserved quantities, \(\epsilon(r)\) denotes the wormhole shape function, and \(\omega(r)\) encodes FD effects. Introducing the impact parameter \(b = L/E\) and linearizing to first order in \(\omega(r)\), the co-rotating and counter-rotating photon orbits take the approximate form
\begin{equation}
b_\pm(r) \simeq r\, e^{-\Phi(r)} \left( 1 \pm \omega(r)\, r\, e^{-\Phi(r)} \right),
\end{equation}
with counter-rotating photons shifted outward and co-rotating photons shifted inward. Cuspy HDE distributions, such as R\'{e}nyi, generate stronger central FD and larger LT splitting, whereas smoother mixed or Moradpour profiles produce nearly circular orbits with smaller asymmetry.

The photon-sphere radius \(r_\mathrm{ph}\) is determined from
\begin{equation*}
\frac{(1 - \omega(r)b)^2}{e^{2\Phi(r)}} = \frac{b^2}{r^2}, \qquad
\frac{d}{dr}\left[\frac{(1 - \omega(r)b)^2}{e^{2\Phi(r)}} - \frac{b^2}{r^2}\right] = 0,
\end{equation*}
and first-order corrections from the redshift function give
\begin{equation*}
r_\mathrm{ph} \simeq r_\mathrm{ph}^{(c)} + \delta r_\mathrm{ph}, \quad
\delta r_\mathrm{ph} \simeq \frac{2 r_\mathrm{ph}^{(c)3} \omega(r_\mathrm{ph}^{(c)})}{2 - r_\mathrm{ph}^{(c)} \Phi'(r_\mathrm{ph}^{(c)})}.
\end{equation*}
where \(r_\mathrm{ph}^{(c)}\) is the photon-sphere radius in the absence of redshift. The induced shifts for the three HDE models are approximately
\begin{eqnarray*}
\delta r_{\mathrm{Sinh}} &\simeq& -\frac{\Phi_0 r_{th}^{3}}{2\left(1-\epsilon'(r_{th})\right)}
e^{-r_{th}/r_c}\left[\frac{\cosh(r_{th}/r_0)}{r_0}-\frac{\sinh(r_{th}/r_0)}{r_c}\right],\nonumber\\
\delta r_{\mathrm{Cosh}} &\simeq& -\frac{\Phi_0 r_{th}^{3}}{2 r_0\left(1-\epsilon'(r_{th})\right)}
\operatorname{sech}\left(\frac{r_{th}}{r_0}\right)\tanh\left(\frac{r_{th}}{r_0}\right),\nonumber\\
\delta r_{\mathrm{Tanh}} &\simeq& -\frac{\Phi_0 r_{th}^{3}}{2 r_0\left(1-\epsilon'(r_{th})\right)}
\operatorname{sech}^{2}\left(\frac{r_{th}}{r_0}\right).
\end{eqnarray*}
Steeper R\'{e}nyi profiles pull the photon sphere inward and create stronger orbit asymmetry, whereas smoother Moradpour profiles induce smaller shifts, keeping photon paths nearly circular.

The apparent boundary of the shadow is expressed as \( b_\pm(r_\mathrm{ph}) \simeq r_\mathrm{ph}\, e^{-\Phi(r_\mathrm{ph})} \left( 1 \pm \omega(r_\mathrm{ph})\, r\, e^{-\Phi(r_\mathrm{ph})} \right)\),
where \(e^{-\Phi(r_\mathrm{ph})}\) sets the overall scale and the LT term introduces asymmetry. R\'{e}nyi HDE enhances shadow size and distortion, while mixed and Moradpour profiles moderate deflection, producing nearly circular shadows. Shadows are most compact and symmetric in the absence of redshift.

Slow rotation shifts co-rotating photons inward and counter-rotating photons outward. This effect is strongest for cuspy R\'{e}nyi distributions and weakest for smoother HDE models. The linearized formulas for \(b_\pm\) and \(r_\mathrm{ph}\) fully capture shadow deformation under the slow-rotation assumption \(|\omega r_\mathrm{ph}| \ll 1\).

The wormhole shape function \(\epsilon(r)\), together with the HDE density, governs throat flaring and the degree of NEC violation. Cuspy distributions require stronger NEC violation and sharper flaring, producing tighter photon trajectories and enhanced lensing. Smoother distributions allow gentler flaring, weaker exotic matter, and wider radial regions, producing larger photon-spheres and more symmetric photon motion. Our results indicate that precise shadow observations could constrain both wormhole rotation and HDE profiles. R\'{e}nyi-supported wormholes produce compact, asymmetric shadows, while mixed and Moradpour-supported wormholes generate larger, nearly circular shadows with reduced FD effects. Photon-sphere shifts, impact parameter asymmetry, and shadow deformation jointly reveal the interplay of wormhole geometry, HDE distribution, and rotation.

In summary, the gradient of \(\Phi(r)\) dictates photon-sphere displacement: cuspy HDE draws it inward, smoother HDE pushes it outward. Slow rotation produces first-order LT splitting, creating shadow asymmetry. R\'{e}nyi profiles require sharper flaring and stronger NEC violation, enhancing deflection and shadow distortion, whereas mixed and Moradpour profiles allow milder flaring, weaker FD, and larger, more symmetric shadows.

\section*{Data Availability Statement} 
The authors declare that the data supporting the findings of this study are available in the article.

\section*{Conflict Of Interest statement} 
The authors declare that they have no conflict of interest.

\section*{Acknowledgements}
This research was funded by the Science Committee of the Ministry of Science and Higher Education of the Republic of Kazakhstan (Grant No. AP23487178).

\bibliography{references}

@article{Einstein:1935tc,
    author = "Einstein, Albert and Rosen, N.",
    title = "{The Particle Problem in the General Theory of Relativity}",
    doi = "10.1103/PhysRev.48.73",
    journal = "Phys. Rev.",
    volume = "48",
    pages = "73--77",
    year = "1935"
}

@article{Gibbons:2015ed,
    author = "Gibbons, Gary W",
    title = "{Editorial note to: Ludwig Flamm, Contributions to Einstein’s theory of gravitation}",
    doi = "",
    journal = "Gen. Relativ. Gravit.",
    volume = "47",
    pages = "71",
    year = "2015"
}

@article{Schwarzschild:1916uq,
    author = "Schwarzschild, Karl",
    title = "{On the gravitational field of a mass point according to Einstein's theory}",
    eprint = "physics/9905030",
    archivePrefix = "arXiv",
    journal = "Sitzungsber. Preuss. Akad. Wiss. Berlin (Math. Phys. )",
    volume = "1916",
    pages = "189--196",
    year = "1916"
}

@article{Schwarzschild:1916ae,
    author = "Schwarzschild, Karl",
    title = "{On the gravitational field of a sphere of incompressible fluid according to Einstein's theory}",
    eprint = "physics/9912033",
    archivePrefix = "arXiv",
    journal = "Sitzungsber. Preuss. Akad. Wiss. Berlin (Math. Phys. )",
    volume = "1916",
    pages = "424--434",
    year = "1916"
}

@article{Flamm:2015ogy,
    author = "Flamm, Ludwig",
    title = "{Republication of: Contributions to Einstein{\textquoteright}s theory of gravitation}",
    doi = "10.1007/s10714-015-1908-2",
    journal = "Gen. Rel. Grav.",
    volume = "47",
    number = "6",
    pages = "72",
    year = "2015"
}

@article{Fuller:1962zza,
    author = "Fuller, Robert W. and Wheeler, John A.",
    title = "{Causality and Multiply Connected Space-Time}",
    doi = "10.1103/PhysRev.128.919",
    journal = "Phys. Rev.",
    volume = "128",
    pages = "919--929",
    year = "1962"
}

@article{Morris:1988cz, 
    author = "Morris, M. S. and Thorne, K. S.",
    title = "{Wormholes in space-time and their use for interstellar travel: A tool for teaching general relativity}",
    doi = "10.1119/1.15620",
    journal = "Am. J. Phys.",
    volume = "56",
    pages = "395--412",
    year = "1988"
}

@article{Morris:1988tu,
    author = "Morris, M. S. and Thorne, K. S. and Yurtsever, U.",
    title = "{Wormholes, Time Machines, and the Weak Energy Condition}",
    doi = "10.1103/PhysRevLett.61.1446",
    journal = "Phys. Rev. Lett.",
    volume = "61",
    pages = "1446--1449",
    year = "1988"
}

@article{Visser:1989kh, 
    author = "Visser, Matt",
    title = "{Traversable wormholes: Some simple examples}",
    eprint = "0809.0907",
    archivePrefix = "arXiv",
    primaryClass = "gr-qc",
    reportNumber = "LA-UR-89-46",
    doi = "10.1103/PhysRevD.39.3182",
    journal = "Phys. Rev. D",
    volume = "39",
    pages = "3182--3184",
    year = "1989"
}

@article{Visser:1989kg,
    author = "Visser, Matt",
    title = "{Traversable wormholes from surgically modified Schwarzschild space-times}",
    eprint = "0809.0927",
    archivePrefix = "arXiv",
    primaryClass = "gr-qc",
    reportNumber = "LA-UR-89-244",
    doi = "10.1016/0550-3213(89)90100-4",
    journal = "Nucl. Phys. B",
    volume = "328",
    pages = "203--212",
    year = "1989"
}

@article{Poisson:1989zz,
    author = "Poisson, Eric and Israel, W.",
    title = "{Inner-horizon instability and mass inflation in black holes}",
    doi = "10.1103/PhysRevLett.63.1663",
    journal = "Phys. Rev. Lett.",
    volume = "63",
    pages = "1663--1666",
    year = "1989"
}

@article{Frolov:1990si,
    author = "Frolov, Valeri P. and Novikov, Igor D.",
    title = "{Physical Effects in Wormholes and Time Machine}",
    reportNumber = "TPI-MINN-90-1-T",
    doi = "10.1103/PhysRevD.42.1057",
    journal = "Phys. Rev. D",
    volume = "42",
    pages = "1057--1065",
    year = "1990"
}

@article{Visser:1990wi,
    author = "Visser, Matt",
    title = "{Quantum wormholes}",
    reportNumber = "PRINT-90-0514 (WASH.U.,ST.LOUIS)",
    doi = "10.1103/PhysRevD.43.402",
    journal = "Phys. Rev. D",
    volume = "43",
    pages = "402--409",
    year = "1991"
}

@article{Martin-Moruno:2013sfa,
    author = "Mart{\'\i}n-Moruno, Prado and Visser, Matt",
    title = "{Classical and quantum flux energy conditions for quantum vacuum states}",
    eprint = "1305.1993",
    archivePrefix = "arXiv",
    primaryClass = "gr-qc",
    doi = "10.1103/PhysRevD.88.061701",
    journal = "Phys. Rev. D",
    volume = "88",
    number = "6",
    pages = "061701",
    year = "2013"
}

@article{Echeverria:1991nk,
    author = "Echeverria, F. and Klinkhammer, G. and Thorne, K. S.",
    title = "{Billiard balls in wormhole space-times with closed timelike curves: Classical theory}",
    doi = "10.1103/PhysRevD.44.1077",
    journal = "Phys. Rev. D",
    volume = "44",
    pages = "1077--1099",
    year = "1991"
}

@article{Hochberg:1997wp,
    author = "Hochberg, David and Visser, Matt",
    title = "{Geometric structure of the generic static traversable wormhole throat}",
    eprint = "gr-qc/9704082",
    archivePrefix = "arXiv",
    doi = "10.1103/PhysRevD.56.4745",
    journal = "Phys. Rev. D",
    volume = "56",
    pages = "4745--4755",
    year = "1997"
}

@article{Willenborg:2018zsv,
    author = "Willenborg, Felix and Grunau, Saskia and Kleihaus, Burkhard and Kunz, Jutta",
    title = "{Geodesic motion around traversable wormholes supported by a massless conformally-coupled scalar field}",
    eprint = "1801.09769",
    archivePrefix = "arXiv",
    primaryClass = "gr-qc",
    doi = "10.1103/PhysRevD.97.124002",
    journal = "Phys. Rev. D",
    volume = "97",
    number = "12",
    pages = "124002",
    year = "2018"
}

@article{Benavides-Gallego:2021lqn,
    author = "Benavides-Gallego, Carlos A. and Han, Wen-Biao and Malafarina, Daniele and Ahmedov, Bobomurat and Abdujabbarov, Ahmadjon",
    title = "{Spinning test particle motion around a traversable wormhole}",
    eprint = "2107.07998",
    archivePrefix = "arXiv",
    primaryClass = "gr-qc",
    doi = "10.1103/PhysRevD.104.084024",
    journal = "Phys. Rev. D",
    volume = "104",
    number = "8",
    pages = "084024",
    year = "2021"
}

@article{Moffat:1991xp, 
    author = "Moffat, J. W. and Svoboda, T.",
    title = "{Traversible wormholes and the negative stress energy problem in the nonsymmetric gravitational theory}",
    doi = "10.1103/PhysRevD.44.429",
    journal = "Phys. Rev. D",
    volume = "44",
    pages = "429--432",
    year = "1991"
}

@article{Bhawal:1992sz,
    author = "Bhawal, Biplab and Kar, Sayan",
    title = "{Lorentzian wormholes in Einstein-Gauss-Bonnet theory}",
    reportNumber = "IITK-92-PHY-09",
    doi = "10.1103/PhysRevD.46.2464",
    journal = "Phys. Rev. D",
    volume = "46",
    pages = "2464--2468",
    year = "1992"
}

@article{Letelier:1993cj,
    author = "Letelier, P. S. and Wang, Anzhong",
    title = "{Spherically symmetric thin shells in Brans-Dicke theory of gravity}",
    doi = "10.1103/PhysRevD.48.631",
    journal = "Phys. Rev. D",
    volume = "48",
    pages = "631--646",
    year = "1993"
}

@article{Vollick:1998qf,
    author = "Vollick, Dan N.",
    title = "{Wormholes in string theory}",
    eprint = "gr-qc/9806096",
    archivePrefix = "arXiv",
    doi = "10.1088/0264-9381/16/5/309",
    journal = "Class. Quant. Grav.",
    volume = "16",
    pages = "1599--1604",
    year = "1999"
}

@article{Myrzakulov:2015kda,
    author = "Myrzakulov, Ratbay and Sebastiani, Lorenzo and Vagnozzi, Sunny and Zerbini, Sergio",
    title = "{Static spherically symmetric solutions in mimetic gravity: rotation curves and wormholes}",
    eprint = "1510.02284",
    archivePrefix = "arXiv",
    primaryClass = "gr-qc",
    doi = "10.1088/0264-9381/33/12/125005",
    journal = "Class. Quant. Grav.",
    volume = "33",
    number = "12",
    pages = "125005",
    year = "2016"
}

@article{Bekenstein:1973ur,
    author = "Bekenstein, Jacob D.",
    title = "{Black holes and entropy}",
    doi = "10.1103/PhysRevD.7.2333",
    journal = "Phys. Rev. D",
    volume = "7",
    pages = "2333--2346",
    year = "1973"
}

@article{Hawking:1975vcx, 
    author = "Hawking, S. W.",
    editor = "Gibbons, G. W. and Hawking, S. W.",
    title = "{Particle Creation by Black Holes}",
    doi = "10.1007/BF02345020",
    journal = "Commun. Math. Phys.",
    volume = "43",
    pages = "199--220",
    year = "1975",
    note = "[Erratum: Commun.Math.Phys. 46, 206 (1976)]"
}

@article{Cohen:1998zx,
    author = "Cohen, Andrew G. and Kaplan, David B. and Nelson, Ann E.",
    title = "{Effective field theory, black holes, and the cosmological constant}",
    eprint = "hep-th/9803132",
    archivePrefix = "arXiv",
    reportNumber = "BUHEP-98-7, DOE-ER-40561-358, INT-98-00-6, UW-PT-97-24",
    doi = "10.1103/PhysRevLett.82.4971",
    journal = "Phys. Rev. Lett.",
    volume = "82",
    pages = "4971--4974",
    year = "1999"
}

@article{Nashed:2026rah,
    author = "Nashed, G. G. L. and Eid, A.",
    title = "{Holographic dark energy as a source for wormholes in modified gravity}",
    doi = "10.1016/j.physletb.2026.140159",
    journal = "Phys. Lett. B",
    volume = "873",
    pages = "140159",
    year = "2026"
}

@article{Tsallis:1987eu, 
    author = "Tsallis, Constantino",
    title = "{Possible Generalization of Boltzmann-Gibbs Statistics}",
    reportNumber = "CBPF-NF-062-87",
    doi = "10.1007/BF01016429",
    journal = "J. Statist. Phys.",
    volume = "52",
    pages = "479--487",
    year = "1988"
}

@article{Moradpour:2017shy,
    author = "Moradpour, H. and Heydarzade, Y. and Darabi, F. and Salako, Ines G.",
    title = "{A Generalization to the Rastall Theory and Cosmic Eras}",
    eprint = "1704.02458",
    archivePrefix = "arXiv",
    primaryClass = "gr-qc",
    doi = "10.1140/epjc/s10052-017-4811-z",
    journal = "Eur. Phys. J. C",
    volume = "77",
    number = "4",
    pages = "259",
    year = "2017"
}

@article{Chaudhary:2025wsd,
    author = "Chaudhary, Sourav and Maurya, S. K. and Kumar, Jitendra and Errehymy, A. and Alessa, Nazek and Abdel-Aty, Abdel -Haleem",
    title = "{A study on the geometric configuration and stability of wormholes in symmetric teleparallel gravity influenced by dark energy and rotational velocity profiles}",
    doi = "10.1140/epjc/s10052-025-14177-y",
    journal = "Eur. Phys. J. C",
    volume = "85",
    number = "5",
    pages = "478",
    year = "2025",
    note = "[Erratum: Eur.Phys.J.C 85, 612 (2025)]"
}

@article{Paul:2025vem,
    author = "Paul, Sat and Maurya, S. K. and Kumar, Jitendra",
    title = "{Stability and existence of wormhole models in F(Q) gravity generated by holographic dark energy densities}",
    doi = "10.1016/j.nuclphysb.2025.116886",
    journal = "Nucl. Phys. B",
    volume = "1014",
    pages = "116886",
    year = "2025"
}

@article{Teo:1998dp,
    author = "Teo, Edward",
    title = "{Rotating traversable wormholes}",
    eprint = "gr-qc/9803098",
    archivePrefix = "arXiv",
    reportNumber = "DAMTP-R-98-17",
    doi = "10.1103/PhysRevD.58.024014",
    journal = "Phys. Rev. D",
    volume = "58",
    pages = "024014",
    year = "1998"
}

@article{Papapetrou:1966,
  author       = {Papapetrou, A.},
  title        = {Champs gravitationnels stationnaires à symétrie axiale},
  journal      = {Ann. de l'I.H.P. Phys. theor.},
  volume       = {A4},
  number       = {2},
  pages        = {83--105},
  year         = {1966},
  url          = {http://eudml.org/doc/75527}
}

@article{Carter:1968a, 
  author       = {Carter, B.},
  title        = {Global Structure of the Kerr Family of Gravitational Fields},
  journal      = {Phys. Rev.},
  volume       = {174},
  number       = {5},
  pages        = {1559},
  year         = {1968},
  doi          = {10.1103/PhysRev.174.1559}
}

@article{Carter:1968b,
  author       = {Carter, B.},
  title        = {Hamilton-Jacobi and Schrödinger Separable Solutions of Einstein's Equations},
  journal      = {Commun. Math. Phys.},
  volume       = {10},
  number       = {4},
  pages        = {280--310},
  year         = {1968},
  doi          = {10.1007/BF03399503}
}

@incollection{Carter:1987,
  author       = {Carter, B.},
  title        = {Mathematical Foundations of the Theory of Relativistic Stellar and Black Hole Configurations},
  booktitle    = {Gravitation in Astrophysics},
  editor       = {Carter, B. and Hartle, J. B.},
  series       = {NATO ASI Series},
  volume       = {156},
  pages        = {63--122},
  year         = {1987},
  doi          = {10.1007/978-1-4613-1897-2_2}
}

@incollection{Roman:1993,
  author       = {Roman, T. A.},
  title        = {Some Thoughts on Energy Conditions and Wormholes},
  booktitle    = {The Tenth Marcel Grossmann Meeting},
  pages        = {1370--1909-1924},
  year         = {2006},
  doi          = {10.1142/9789812704030_0236}
}

@article{Kar:1994,
  author       = {Kar, S. and Sahdev, D.},
  title        = {Evolving Lorentzian wormholes},
  journal      = {Phys. Rev. D},
  volume       = {53},
  number       = {2},
  pages        = {722--731},
  year         = {1996},
  doi          = {10.1103/PhysRevD.53.722}
}

@article{Kar:1996,
  author       = {Kar, S.},
  title        = {Evolving Lorentzian wormholes and energy conditions},
  journal      = {Phys. Rev. D},
  volume       = {49},
  number       = {2},
  pages        = {862--865},
  year         = {1996},
  doi          = {10.1103/PhysRevD.49.862}
}

@article{Butcher:2015sea,
  author       = {Butcher, L. M.},
  title        = {Traversable Wormholes and Classical Scalar Fields},
  journal      = {Phys. Rev. D},
  volume       = {91},
  number       = {12},
  pages        = {124031},
  year         = {2015},
  doi          = {10.1103/PhysRevD.91.124031}
}

@article{Hartle:1967ha,
    author = "Hartle, James B and Sharp, David H",
    title = "{Variational principle for the equilibrium of a relativistic, rotating star}",
    doi = "10.1086/149002",
    journal = "Astrophys. J.",
    volume = "147",
    pages = "317",
    year = "1967"
}

@article{Hartle:1967he,
    author = "Hartle, James B.",
    title = "{Slowly rotating relativistic stars. 1. Equations of structure}",
    doi = "10.1086/149400",
    journal = "Astrophys. J.",
    volume = "150",
    pages = "1005--1029",
    year = "1967"
}

@article{Li:2004rb,
    author = "Li, Miao",
    title = "{A Model of holographic dark energy}",
    eprint = "hep-th/0403127",
    archivePrefix = "arXiv",
    doi = "10.1016/j.physletb.2004.10.014",
    journal = "Phys. Lett. B",
    volume = "603",
    pages = "1",
    year = "2004"
}

@article{Hu:2006ar,
    author = "Hu, Bo and Ling, Yi",
    title = "{Interacting dark energy, holographic principle and coincidence problem}",
    eprint = "hep-th/0601093",
    archivePrefix = "arXiv",
    doi = "10.1103/PhysRevD.73.123510",
    journal = "Phys. Rev. D",
    volume = "73",
    pages = "123510",
    year = "2006"
}

@article{Myung:2004ch,
    author = "Myung, Yun Soo",
    title = "{Holographic principle and dark energy}",
    eprint = "hep-th/0412224",
    archivePrefix = "arXiv",
    reportNumber = "INJE-TP-04-08",
    doi = "10.1016/j.physletb.2005.02.006",
    journal = "Phys. Lett. B",
    volume = "610",
    pages = "18--22",
    year = "2005"
}

@article{Moradpour:2017fmq,
    author = "Moradpour, H. and Sheykhi, A. and Corda, C. and Salako, Ines G.",
    title = "{Implications of the generalized entropy formalisms on the Newtonian gravity and dynamics}",
    eprint = "1711.10336",
    archivePrefix = "arXiv",
    primaryClass = "physics.gen-ph",
    doi = "10.1016/j.physletb.2018.06.040",
    journal = "Phys. Lett. B",
    volume = "783",
    pages = "82--85",
    year = "2018"
}

@article{Manoharan:2022qll,
    author = "Manoharan, Manosh T. and Shaji, N. and Mathew, Titus K.",
    title = "{Holographic dark energy from the laws of thermodynamics with R{\'e}nyi entropy}",
    eprint = "2208.08736",
    archivePrefix = "arXiv",
    primaryClass = "gr-qc",
    doi = "10.1140/epjc/s10052-023-11202-w",
    journal = "Eur. Phys. J. C",
    volume = "83",
    number = "1",
    pages = "19",
    year = "2023"
}

@article{Mamon:2016wow,
    author = "Mamon, Abdulla Al and Bamba, Kazuharu and Das, Sudipta",
    title = "{Constraints on reconstructed dark energy model from SN Ia and BAO/CMB observations}",
    eprint = "1607.06631",
    archivePrefix = "arXiv",
    primaryClass = "gr-qc",
    doi = "10.1140/epjc/s10052-016-4590-y",
    journal = "Eur. Phys. J. C",
    volume = "77",
    number = "1",
    pages = "29",
    year = "2017"
}

@article{Anderhalden:2012qt,
    author = "Anderhalden, Donnino and Diemand, Juerg and Bertone, Gianfranco and Maccio, Andrea V. and Schneider, Aurel",
    title = "{The Galactic Halo in Mixed Dark Matter Cosmologies}",
    eprint = "1206.3788",
    archivePrefix = "arXiv",
    primaryClass = "astro-ph.CO",
    doi = "10.1088/1475-7516/2012/10/047",
    journal = "JCAP",
    volume = "10",
    pages = "047",
    year = "2012"
}

@article{Garattini:2023wgk,
    author = "Garattini, Remo and Channuie, Phongpichit",
    title = "{Traversable wormholes supported by holographic dark energy with a modified equation of state}",
    eprint = "2311.04620",
    archivePrefix = "arXiv",
    primaryClass = "gr-qc",
    doi = "10.1016/j.nuclphysb.2024.116589",
    journal = "Nucl. Phys. B",
    volume = "1005",
    pages = "116589",
    year = "2024"
}

@article{Moradpour:2018ivi,
    author = "Moradpour, H. and Moosavi, S. A. and Lobo, I. P. and Morais Gra{\c{c}}a, J. P. and Jawad, A. and Salako, I. G.",
    title = "{Thermodynamic approach to holographic dark energy and the R{\'e}nyi entropy}",
    eprint = "1803.02195",
    archivePrefix = "arXiv",
    primaryClass = "physics.gen-ph",
    doi = "10.1140/epjc/s10052-018-6309-8",
    journal = "Eur. Phys. J. C",
    volume = "78",
    number = "10",
    pages = "829",
    year = "2018"
}

\end{document}